# Title: Spin Photovoltaic Effect in Magnetic van der Waals Heterostructures

**Authors:** Tiancheng Song[1], Eric Anderson[1], Matisse Wei-Yuan Tu[2], Kyle Seyler[1], Takashi Taniguchi[3], Kenji Watanabe[4], Michael A. McGuire[5], Xiaosong Li[6], Ting Cao[6], Di Xiao[7], Wang Yao[8], Xiaodong Xu[1, 6]*

**Affiliations:**
[1]Department of Physics, University of Washington, Seattle, Washington 98195, USA.
[2]Department of Physics, National Cheng Kung University, Tainan 70101, Taiwan.
[3]International Center for Materials Nanoarchitectonics, National Institute for Materials Science, 1-1 Namiki, Tsukuba 305-0044, Japan.
[4]Research Center for Functional Materials, National Institute for Materials Science, 1-1 Namiki, Tsukuba 305-0044, Japan.
[5]Materials Science and Technology Division, Oak Ridge National Laboratory, Oak Ridge, Tennessee 37831, USA.
[6]Department of Materials Science and Engineering, University of Washington, Seattle, Washington 98195, USA.
[7]Department of Physics, Carnegie Mellon University, Pittsburgh, Pennsylvania 15213, USA.
[8]Department of Physics, University of Hong Kong, Hong Kong, China.

*Correspondence to: xuxd@uw.edu

**Abstract:**

**The development of van der Waals (vdW) crystals and their heterostructures has created a fascinating platform for exploring optoelectronic properties in the two-dimensional (2D) limit. With the recent discovery of 2D magnets, the control of the spin degree of freedom can be integrated to realize 2D spin-optoelectronics with spontaneous time-reversal symmetry breaking. Here, we report spin photovoltaic effects in vdW heterostructures of atomically thin magnet chromium triiodide ($CrI_3$) sandwiched by graphene contacts. In the absence of a magnetic field, the photocurrent displays a distinct dependence on light helicity, which can be tuned by varying the magnetic states and photon energy. Circular polarization-resolved absorption measurements reveal that these observations originate from magnetic-order-coupled and thus helicity-dependent charge-transfer exciton states. The photocurrent displays multiple plateaus as the magnetic field is swept, which are associated with different spin configurations enabled by the layered antiferromagnetism and spin-flip transitions in $CrI_3$. Remarkably, giant photo-magnetocurrent is observed, which tends to infinity for a small applied bias. Our results pave the way to explore emergent photo-spintronics by engineering magnetic vdW heterostructures.**

**Main text:**

Spintronics aims at manipulating the spin degree of freedom in electronic systems for novel functionalities[1]. In optoelectronics, the generation and control of spins can open up emerging opportunities for spin-optoelectronics, enabling the exploration of new spin photovoltaic effects and spin photocurrents. In various magnetic heterostructures, spin photovoltaic effects can be realized by different mechanisms. For instance, a spin voltage arises from spin-dependent excitation at the interface of a nonmagnetic metal in close proximity to a magnetic insulator[2]. In spin valves and magnetic p-n junctions, spin injection and accumulation can be induced by the spin-dependent injection process of the photogenerated carriers at the interfaces with ferromagnetic contacts[3–6]. Alternatively, in materials without intrinsic magnetic order, circularly polarized light can generate spin photocurrents via the circular photogalvanic effect[7–10]. Among these materials and heterostructures, two-dimensional (2D) materials, in particular, transition metal dichalcogenides (TMDs) have established themselves as a promising system for spin-optoelectronics due to their spin-valley dependent properties and enhanced photoresponsivity from strong excitonic effects[9–17].

The recent discovery of 2D vdW magnets provide a new platform for spin photovoltaic effects based on atomically thin materials with intrinsic magnetic order[18–21]. Among these magnets, chromium triiodide ($CrI_3$) is particularly interesting because of its layered antiferromagnetism (AFM), where the ferromagnetic monolayers with out-of-plane magnetizations are antiferromagnetically coupled to each other, as shown in Fig. 1a. The spin configurations can be manipulated by an external magnetic field which switches the sample between the AFM ground states and the fully spin-polarized states via a series of spin-flip transitions[18]. Multiple magnetic states become accessible as the number of $CrI_3$ layers increases, possibly enabling multiple states of the resulting spin photocurrent[22]. Moreover, given the reported strong magneto-optical and excitonic effects[23,24], atomically thin $CrI_3$ should provide an ideal platform to explore spin-optoelectronic effects in the atomically thin limit[21].

**Photocurrent response of $CrI_3$ junction device**

To investigate photocurrent response from $CrI_3$, a vertical heterostructure was fabricated for efficient photodetection. All measurements were carried out at a temperature of 2K with a magnetic field in the out-of-plane direction and a linearly polarized laser excitation, unless otherwise specified. As shown in Fig. 1a, the heterostructure consists of an atomically thin $CrI_3$ flake sandwiched by two graphene sheets as bias electrodes, encapsulated by thin hexagonal boron nitride (hBN) to avoid degradation (see Methods). Such a structure is essentially the same as a magnetic tunnel junction (MTJ), which has been used to realize large tunneling magnetoresistance via the spin-filtering effect enabled by the layered antiferromagnetism in $CrI_3$[22,25–27]. Using a four-layer $CrI_3$ device (D2) as an example, without optical illumination, the current-bias characteristics (*I-V* curves) of the device behave as a typical tunnel junction. The tunneling current is suppressed in the low bias regime and dominated by Fowler-Nordheim tunneling at high bias[22,25–27] (black curve, Fig. 1b).

Compared to the dark condition case, a substantial enhancement of the current is observed with photoexcitation of carriers in the low bias regime. The red curve in Fig. 1b is obtained with 1.96 eV (632.8 nm) laser excitation focused to a ~1 µm spot size at normal incidence, with an optical power of 1 µW. This carrier collection process is highly efficient in the vertical junction structure of atomically thin $CrI_3$, due to the reduced requirement of the carrier diffusion length. At zero bias, a net photocurrent $I_{ph}$ is also generated (inset, Fig. 1b). This zero bias photocurrent can be attributed to the asymmetric potential of the junction[11,12], which could originate from the potential difference between the top and bottom graphene/$CrI_3$ interfaces. Applying a bias voltage induces an external electric field, which can modulate the magnitude as well as reverse the direction of the photocurrent. When the applied bias compensates the built-in electric field such that the net current is zero, the system becomes equivalent to an open circuit, allowing us to measure the photogenerated open-circuit voltage ($V_{oc}$).

We investigate the spatial distribution of the photocurrent by employing scanning photocurrent microscopy. Figure 1d shows the optical microscopy image of a trilayer $CrI_3$ device (D1), which has a large junction area, with the corresponding photocurrent map taken at zero bias, shown in Fig. 1e. We also employ reflective magnetic circular dichroism (RMCD) microscopy to map out the trilayer $CrI_3$ flake shown in Fig. 1f, which measures the out-of-plane magnetization of the device at zero field (see Methods). By comparing the photocurrent map with the microscopy image and the RMCD map, the photoactive region can be identified at the junction region where the top and bottom graphene electrodes overlap.

Figure 1c shows the photon energy dependence of the photocurrent. $I_{ph}$ increases sharply when the photon energy exceeds 1.7 eV. By comparison to the differential reflectance ($\Delta R/R$) measurement of a trilayer $CrI_3$ on a sapphire substrate (see Methods), we attribute the strong photocurrent response to the optical excitation of ligand-to-metal charge-transfer excitons[23,24]. We do not observe photocurrent enhancement corresponding to the excitation of the lower energy exciton at 1.5 eV. This is possibly due to its larger binding energy and more localized nature than the charge transfer excitons[24]. Notably, the photoresponsivity reaches 10 mA $W^{-1}$, which is already comparable to that achieved in the devices based on TMD semiconductors under similar conditions[11,12] (the photocurrent map in Supplementary Fig. S3 shows the photoresponsivity reaches 10 mA $W^{-1}$, and Supplementary Fig. S1 shows the laser excitation power dependence of photocurrent).

**Magnetic order dependence of photocurrent**

The photocurrent response has a strong dependence on the magnetic order. Figure 2a shows zero bias $I_{ph}$ as a function of the external magnetic field ($\mu_0 H$) in a four-layer $CrI_3$ with an optical power of 1 µW. As $\mu_0 H$ is swept, $I_{ph}$ exhibits several sharp transitions and multiple plateaus. Figure 2b shows RMCD signal with the corresponding magnetic states labeled, as identified in the previous studies[22,28]. For simplicity, only the positive magnetic field side is shown. The full field data with magnetic states assignment can be found in Supplementary Fig. S2. Comparison of Figs. 2a and b shows that the multiple photocurrent plateaus are associated with the distinct magnetic states. The low and high photocurrent plateaus at low and high fields can be assigned to the AFM ground states and fully spin-polarized states, respectively.

Interestingly, the intermediate magnetic states (either ↑↓↑↑ or ↑↑↓↑) result in a lower photocurrent than the AFM ground states. This non-monotonic photocurrent response to the magnetic states is distinct from the monotonic increase of the tunneling conductance due to the spin-filtering effect[22,29]. As a comparison, Fig. 2c shows the tunneling current of the same device measured as a function of $\mu_0 H$ at 80 mV bias under dark condition. In sharp contrast to $I_{ph}$, the tunneling current increases monotonically and dramatically as the spins in each layer are aligned from ↑↓↑↓ to ↑↑↑↑, because the current-blocking antiparallel interfaces are removed. Note that the tunneling current varies by two orders of magnitude for different magnetic states[22,25–27], while there is only a two-fold difference in the photocurrent. For tunneling under dark condition, the electron energy is below the $CrI_3$ conduction bands. In contrast, the optical excitation generates photoexcited carriers in the conduction bands, and their asymmetric extraction by the top and bottom graphene electrodes results in the measured photocurrent. The spin configurations of $CrI_3$ determine the layer distribution of the wavefunction of photoexcited carriers, through which the extraction efficiencies at the top and bottom electrodes can be affected, accounting for the non-monotonic magnetic state dependence of photocurrent[11,12,30,31] (see Supplementary Text S1). A thorough understanding of the magnetic-state-dependent exciton formation and dissociation processes and the photocurrent generation mechanism in a few-layer $CrI_3$ requires future theoretical and experimental studies.

In analogy to giant magnetoresistance and tunnel magnetoresistance[1,32–35], which are of great importance for spintronics, our spin-optoelectronic device exhibits a novel photo-magnetocurrent effect[3]. Figure 2d shows the $I_{ph}$-$V$ curves corresponding to the AFM ground state (0 T, black curve) and fully spin-polarized state (2.5 T, red curve), respectively. For the short-circuit condition, the fully spin-polarized state generates a higher photocurrent, whereas the AFM ground state gives a larger open-circuit voltage magnitude. To quantify this magnetic state dependence, we define photo-magnetocurrent ratio as $MC_{ph} = \frac{I_{ph}^{p} - I_{ph}^{ap}}{I_{ph}^{ap}}$, where $I_{ph}^{p}$ and $I_{ph}^{ap}$ are the photocurrents corresponding to the fully spin-polarized state (parallel) and AFM ground state (antiparallel). Figure 2e shows the absolute value of $MC_{ph}$ as a function of bias extracted from the $I_{ph}$-$V$ curves in Fig. 2d. Remarkably, a giant $MC_{ph}$ is observed within a range of bias voltage indicated by the red shading. This can be attributed to the magnetic-state-dependent open-circuit voltage, where at certain bias $I_{ph}^{ap}$ goes to zero while $I_{ph}^{p}$ is still finite, leading to a giant $MC_{ph}$ ratio tending to infinity[3]. This demonstrates a proof-of-concept photo-modulated magnetocurrent effect. Achieving such a giant and tunable photo-magnetocurrent could be useful for optically driven magnetic sensing and data storage technologies.

**Dependence of photocurrent on light helicity**

The broken time-reversal symmetry of our system should also enable a light helicity dependence of the spin photocurrent. Here, we use the trilayer $CrI_3$ device (D1) with 1.96 eV excitation as an example. The magnetization is set in the fully spin-polarized state, either ↑↑↑ (2 T) or ↓↓↓ (-2 T). As the light helicity is switched between $\sigma^+$ and $\sigma^-$, the photocurrent exhibits a clear circular polarization dependence. As shown in Fig. 3a, the ↑↑↑ state (red dots) gives a higher photocurrent for photon helicity $\sigma^-$ (135°) than $\sigma^+$ (45°). In contrast, the ↓↓↓ state (black dots) exhibits the exact opposite helicity dependence, consistent with the time-reversal operation that connects the two fully spin-polarized states.

We define the difference of photocurrent between $\sigma^+$ and $\sigma^-$ excitation as $\Delta I_{ph}\,[\sigma^+\text{-}\sigma^-] = I_{ph}(\sigma^+) - I_{ph}(\sigma^-)$. The degree of helicity is then denoted by $\Delta I_{ph}\,[\sigma^+\text{-}\sigma^-]/(I_{ph}(\sigma^+)+I_{ph}(\sigma^-))$. To fully understand the interplay between the helicity-dependent photocurrent and the underlying magnetic order, we measure $\Delta I_{ph}\,[\sigma^+\text{-}\sigma^-]$ and the degree of helicity as a function of $\mu_0 H$ shown in Fig. 3b (see Methods). Four distinct plateaus are observed, which behave essentially the same as the RMCD signal versus $\mu_0 H$ measured from the same device with the same 1.96 eV laser (Fig. 3c). We can thus assign the corresponding magnetic states to each $\Delta I_{ph}\,[\sigma^+\text{-}\sigma^-]$ plateau. There is notable magnetic hysteresis of $\Delta I_{ph}\,[\sigma^+\text{-}\sigma^-]$ centered at zero field due to switching between the ↑↓↑ and ↓↑↓ AFM coupled ground states. In addition, the four-layer CrI$_3$ device (D2) shows a similar magnetic-state-coupled helicity dependence of the photocurrent (Supplementary Fig. S2).

**Optical selection rules of magnetic-order-coupled charge-transfer excitons**

We find that this unique spin photovoltaic effect originates from the helicity dependence of charge-transfer excitons in CrI$_3$, which couple to the underlying magnetic order. Figure 4a shows circular polarization-resolved differential reflectance ($\Delta R/R$) measurements of a trilayer CrI$_3$ on a sapphire substrate. Data from all four magnetic states, {↑↑↑ (2 T), ↓↓↓ (-2 T), ↑↓↑ (0 T), ↓↑↓ (0 T)}, are shown. Evidently, the $\sigma^+/\sigma^-$ (red/blue dots) absorption peaks diverge in both energy and intensity, and are determined by the magnetic state. This observation is consistent with the magnetic-order-coupled charge-transfer excitons calculated by the many-body perturbation theory[24]. The helicity-dependent absorption reveals the optical selection rules of the charge-transfer transitions between the spin-polarized valence and conduction bands[24,36], and thus result in the observed helicity-dependent spin photovoltaic effect.

Starting with the ↑↑↑ state, the 1.96 eV (632.8 nm) excitation indicated by the red dashed line is near the resonance of the $\sigma^-$ polarized charge-transfer exciton, while the $\sigma^+$ resonance is at 2.07 eV, about 110 meV higher. The stronger absorption for $\sigma^-$ than $\sigma^+$ results in higher photocurrent for the $\sigma^-$ excitation vs $\sigma^+$, as shown in Fig. 3a. For the ↓↓↓ state, the absorption peaks are switched between $\sigma^+$ and $\sigma^-$ compared to the ↑↑↑ state, which agrees with the opposite helicity dependence of the ↓↓↓ state photocurrent (Fig. 3a). The magnetic ground states at zero magnetic field, ↑↓↑ and ↓↑↓, also give notable but opposite divergence between the $\sigma^+$ and $\sigma^-$ absorption peaks, due to their opposite net magnetizations. This divergence between $\sigma^+$ and $\sigma^-$ vanishes above the critical temperature of trilayer CrI$_3$ (Supplementary Fig. S4). Note that for even number layers, the vanishing net magnetization at the AFM ground states leads to vanishing helicity dependence of the charge transfer excitons (see Supplementary Fig. S5 for the $\Delta R/R$ measurement of a six-layer CrI$_3$ on a sapphire substrate). All these observations confirm the underlying magnetic order as the origin of the helicity dependence of the charge-transfer excitons.

The circularly polarized optical selection rules of charge-transfer excitons also enable the control of photocurrent helicity dependence by tuning the optical excitation energy. We choose three selected photon energies, indicated by the dashed lines in Fig. 4a for the magnetic state ↓↓↓ panel. These three energies represent stronger $\sigma^+$ absorption than $\sigma^-$ (1.88 eV), nearly equal absorption (2.01 eV), and stronger $\sigma^-$ absorption than $\sigma^+$ (2.13 eV), respectively. Figure 4b shows the corresponding helicity-dependent photocurrent at these photon energies. For the ↓↓↓ state, the $\sigma^+$ excitation at 1.88 eV gives a higher photocurrent than $\sigma^-$ excitation, and this scenario is reversed for 2.13 eV excitation. The helicity dependence nearly vanishes for the 2.01 eV excitation, consistent with the observed equal absorption of $\sigma^+$ and $\sigma^-$ polarized light. Figure 4c shows $\Delta I_{ph}$ [$\sigma^+$-$\sigma^-$] at several photon energies. Clearly, the helicity dependence exhibits a change in sign around 2.01 eV. This $\Delta I_{ph}$ [$\sigma^+$-$\sigma^-$] as a function of photon energy matches well with the overlaid $\Delta R/R$ helicity difference ($\Delta R/R(\sigma^+)$-$\Delta R/R(\sigma^-)$). As expected, the photocurrent behavior of the magnetic state ↑↑↑, the red dots and curves in Fig. 4b and c, is the time reversal of ↓↓↓ state.

**Conclusions**

We explore spin photovoltaic effects in atomically thin $CrI_3$ vdW heterostructures. The photocurrent exhibits distinct responses to the spin configurations in $CrI_3$, together with a giant photo-magnetocurrent effect. The combination of our helicity-dependent photocurrent and circular polarization-resolved absorption measurements reveals the emergent interplay between the spin photocurrent and the underlying excitons, intrinsic magnetic order, photon energy and helicity. Our work unlocks the spin degree of freedom in 2D optoelectronics with intrinsic magnetism, establishing atomically thin $CrI_3$ as a new platform for exploring emergent spin-optoelectronic phenomena and designing magneto-optoelectronic devices with circular polarization-resolved capability.

**Data availability:** The data that support the findings of this study are available from the corresponding author upon reasonable request.

**Acknowledgments:** This work was mainly supported by Air Force Office of Scientific Research (AFOSR) Multidisciplinary University Research Initiative (MURI) program, grant no. FA9550-19-1-0390. The tunneling current measurements were partially supported by NSF-DMR-1708419. The magnetic circular dichroism and optical spectroscopy measurements were partially supported by the Department of Energy, Basic Energy Sciences, Materials Sciences and Engineering Division (DE-SC0018171). Theoretical understanding is partially supported by NSF MRSEC DMR-1719797, and RGC of HKSAR (17303518). M.A.M. was supported by the US Department of Energy, Office of Science, Basic Energy Sciences, Materials Sciences and Engineering Division. K.W. and T.T. acknowledge support from the Elemental Strategy Initiative conducted by the MEXT, Japan, Grant Number JPMXP0112101001, JSPS KAKENHI Grant Number JP20H00354 and the CREST (JPMJCR15F3), JST. X.X. acknowledges the support from the State of Washington funded Clean Energy Institute and the Boeing Distinguished Professorship in Physics.

**Author contributions:** X.X., T.S. conceived the experiment. T.S. fabricated the devices and performed the measurements, assisted by E.A. K.S. performed a preliminary reflection measurement. T.S., E.A., X.X. analyzed and interpreted the results. M.W.-Y.T. and W.Y. provided theory support, with input from T.C. and D.X., and X.L.. T.T. and K.W. synthesized the hBN crystals. M.A.M. synthesized and characterized the bulk CrI$_3$ crystals. T.S., E.A., X.X. wrote the paper with inputs from all the authors. All the authors discussed the results.

**Competing interests:** The authors declare no competing interests.

**Methods:**

**Device fabrication:** The monolayer/few-layer graphene and 15-25 nm hBN flakes were mechanically exfoliated onto either 285 nm or 90 nm SiO$_2$/Si substrates and examined by optical and atomic force microscopy under ambient conditions. Only atomically clean and smooth flakes were used for making devices. V/Au (5/50 nm) metal electrodes were deposited onto a 285 nm SiO2/Si substrate using standard electron beam lithography with a bilayer resist (A4 495 and A4 950 poly (methyl methacrylate) (PMMA)) and electron beam evaporation. CrI$_3$ crystals were exfoliated onto 90 nm SiO$_2$/Si substrates in an inert gas glovebox with water and oxygen concentration less than 0.1 ppm. The CrI$_3$ flake thickness was identified by optical contrast with respect to the substrate using the established optical contrast model[18]. The layer assembly was performed in the glovebox using a polymer-based dry transfer technique. The flakes were picked up sequentially: top hBN, top graphene contact, $CrI_3$, bottom graphene contact, bottom hBN. The resulting stacks were then transferred and released on the pre-patterned electrodes. In the resulting heterostructure, the $CrI_3$ flake is fully encapsulated on both sides, and the top/bottom graphene flakes are connected to the pre-patterned electrodes. Finally, the polymer was dissolved in chloroform for less than one minute to minimize the sample exposure to ambient conditions.

**Photocurrent measurement:** The photocurrent measurements were performed in a closed-cycle cryostat (attoDRY 2100) at a temperature of 2 K and an out-of-plane magnetic field up to 9 T. A 632.8 nm HeNe laser was focused to a ~1 µm spot size at normal incidence to generate photocurrent. Figure 1a shows the schematic of $CrI_3$ junction devices. For DC measurement, a bias voltage ($V$) was applied to the top graphene contact with the bottom contact grounded. The resulting photocurrent ($I_{ph}$) or tunneling current ($I_t$) was amplified and measured by a current preamplifier (DL Instruments; Model 1211). For AC measurement, a standard lock-in technique was used to measure the change in photocurrent with Stanford Research Systems SR830. For the photon energy dependence measurement, a SolsTiS continuous-wave widely tunable laser was used to generate photocurrent. For the photon helicity dependence measurement, a motorized precision rotation mount was used to rotate an achromatic quarter-wave plate with respect to the linear polarized incident laser beam.

**Optical measurements:** The reflective magnetic circular dichroism (RMCD) and Kerr rotation measurements were performed in two similar cryostats (attoDRY 2100 and Quantum Design OptiCool) under the same experimental conditions. A 632.8 nm HeNe laser was used to probe the device at normal incidence with a fixed power of 1 uW. The AC lock-in measurement technique used to measure the RMCD and Kerr rotation signal closely follows the previous RMCD and magneto-optical Kerr effect (MOKE) measurements of the magnetic order in atomically-thin $CrI_3$[18,22]. For the differential reflectance measurements, we spatially filtered a tungsten halogen lamp and focused the beam to a ~3 µm spot size on the $CrI_3$. The reflected light was deflected with a beamsplitter and detected by a spectrometer and a liquid-nitrogen-cooled charge-coupled device, which enabled signal measurement from 1.4 eV to 3 eV. To obtain the differential reflectance, we subtracted and normalized the $CrI_3$ reflectance by the reflectance of the sapphire substrate. The absorbance of $CrI_3$ is proportional to the differential reflectance, which can be determined as $\frac{1}{4}(n^2-1)\Delta R/R$ [23,37].

**Main figures:**

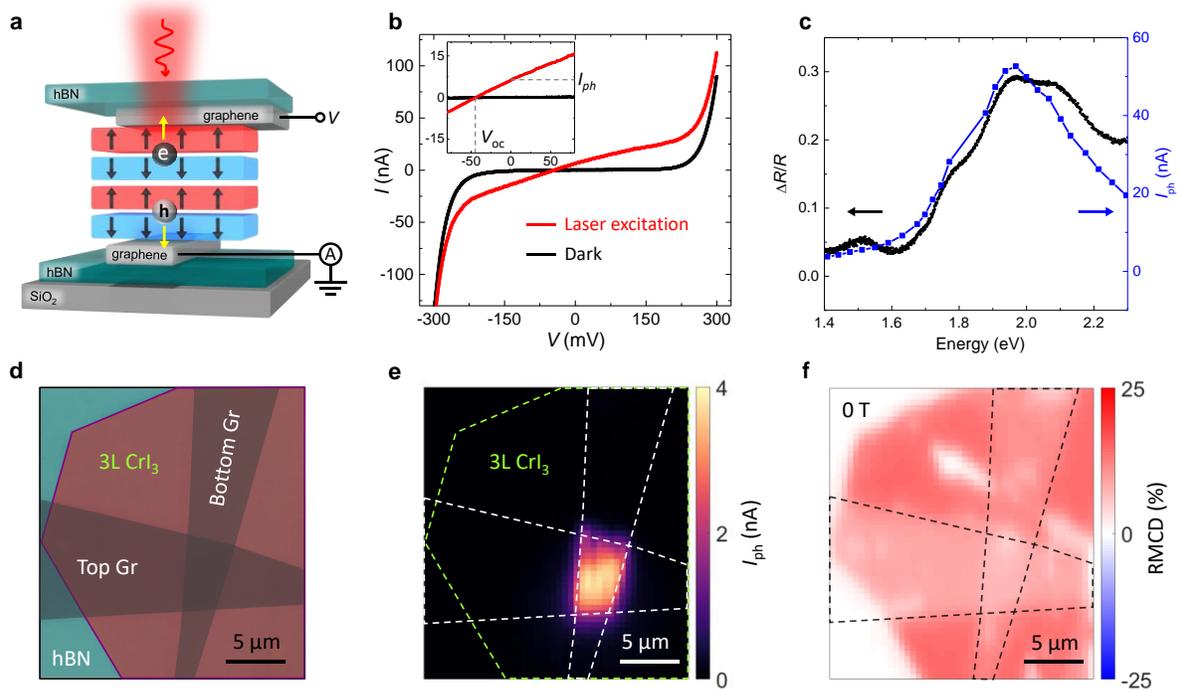

**Figure 1 | Photocurrent response of CrI$_3$ junction device. a**, Schematic of a four-layer CrI$_3$ junction device in AFM ground state (↑↓↑↓), with top and bottom graphene contacts and hBN encapsulation. **b**, *I-V* curves of a four-layer CrI$_3$ junction (D2) under dark condition (black curve) and with 1µW of 1.96 eV laser excitation (red curve). Inset is a zoomed-in view of generated photocurrent at zero bias $I_{ph}$ and open circuit voltage $V_{OC}$. **c**, Differential reflectance ($\Delta R/R$, black dots) and photocurrent ($I_{ph}$, blue squares) as a function of photon energy for trilayer CrI$_3$ at -2 T. The photocurrent is measured from a trilayer CrI$_3$ junction device (D1) with an optical power of 10 µW. **d**, Optical microscopy image of the 3L CrI$_3$ junction device (D1) (scale bar, 5 µm). **e**, **f**, Spatial maps of photocurrent and RMCD signal measured from the same device at 0 T with an optical power of 1 µW (scale bar, 5 µm).

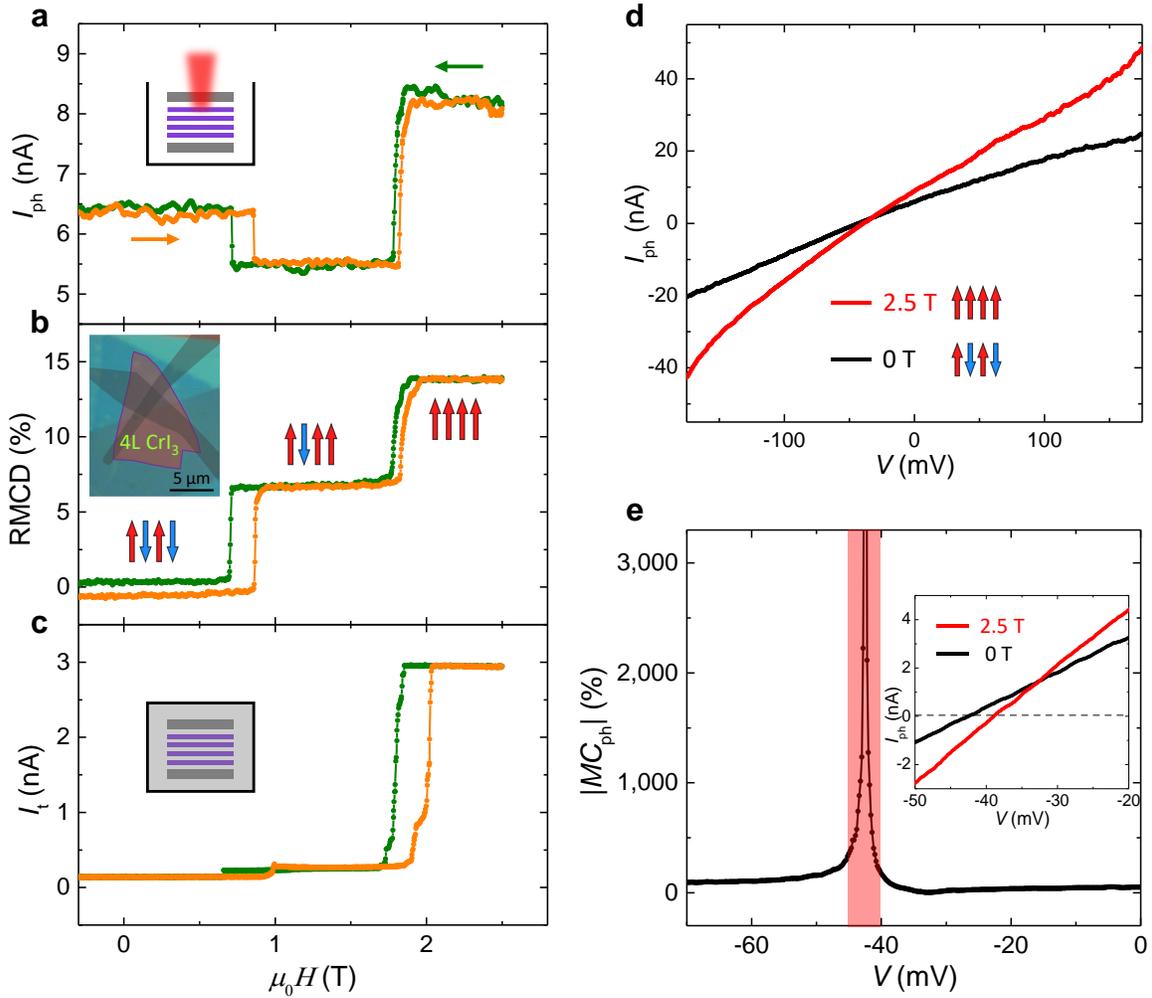

**Figure 2 | Dependence of photocurrent on magnetic order of four-layer CrI$_3$. a**, Photocurrent as a function of external magnetic field ($\mu_0H$) measured from the four-layer CrI$_3$ junction device (D2) with an optical power of 1 µW. Green (orange) curve corresponds to decreasing (increasing) magnetic field. **b**, RMCD as a function of $\mu_0H$ for the same device. Insets show the corresponding magnetic states and the optical microscopy image of the device (D2). **c**, Tunneling current ($I_t$) as a function of $\mu_0H$ measured from the same device at 80 mV bias under dark condition. Insets are schematics of the device with laser excitation and under dark condition. **d**, $I_{ph}$-$V$ curves for the four-layer CrI$_3$ in the AFM ground state (↑↓↑↓, 0 T, black curve) and the fully spin-polarized state (↑↑↑↑, 2.5 T, red curve). **e**, Magnitude of the photo-magnetocurrent ratio as a function of bias extracted from the $I_{ph}$-$V$ curves in **d**. The red shading denotes the bias range where |$MC_{ph}$| tends to infinity. Inset is a zoomed-in view of the $I_{ph}$-$V$ curves in **d**.

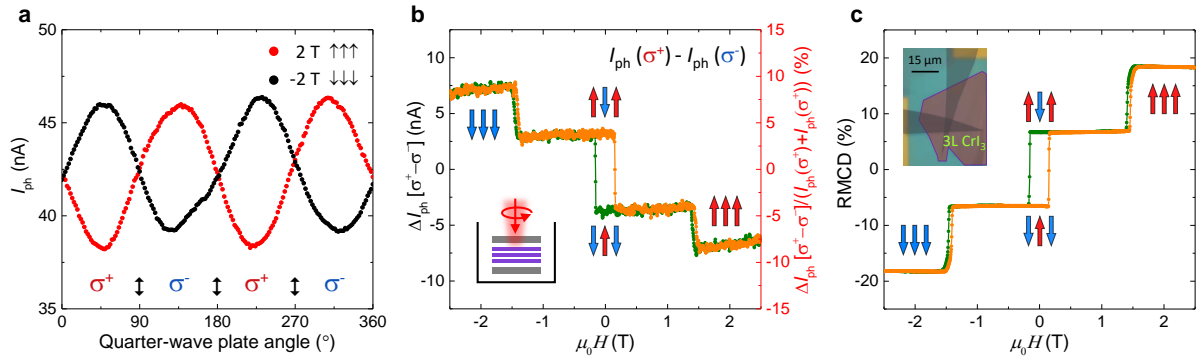

**Figure 3 | Helicity dependence of photocurrent in trilayer CrI$_3$. a**, Photocurrent as a function of quarter-wave plate angle for ↑↑↑ state (2 T, red dots) and ↓↓↓ state (-2 T, black dots) measured from the trilayer CrI$_3$ junction device (D1) with an optical power of 10 µW. Vertical arrows represent linearly polarized light. **b**, The change in photocurrent ($\Delta I_{ph}$ [$\sigma^+$-$\sigma^-$] = $I_{ph}(\sigma^+)$-$I_{ph}(\sigma^-)$) as a function of $\mu_0 H$ measured from the same device with an optical power of 10 µW. The degree of helicity $\Delta I_{ph}$ [$\sigma^+$-$\sigma^-$]/(($I_{ph}(\sigma^+)$+$I_{ph}(\sigma^-)$)) given on right axis. Insets show the corresponding magnetic states and schematic of the device with circularly polarized light excitation. **c**, RMCD as a function of $\mu_0 H$ for the same device. Insets show the corresponding magnetic states and the optical microscopy image of the device (D1) (scale bar, 15 µm).

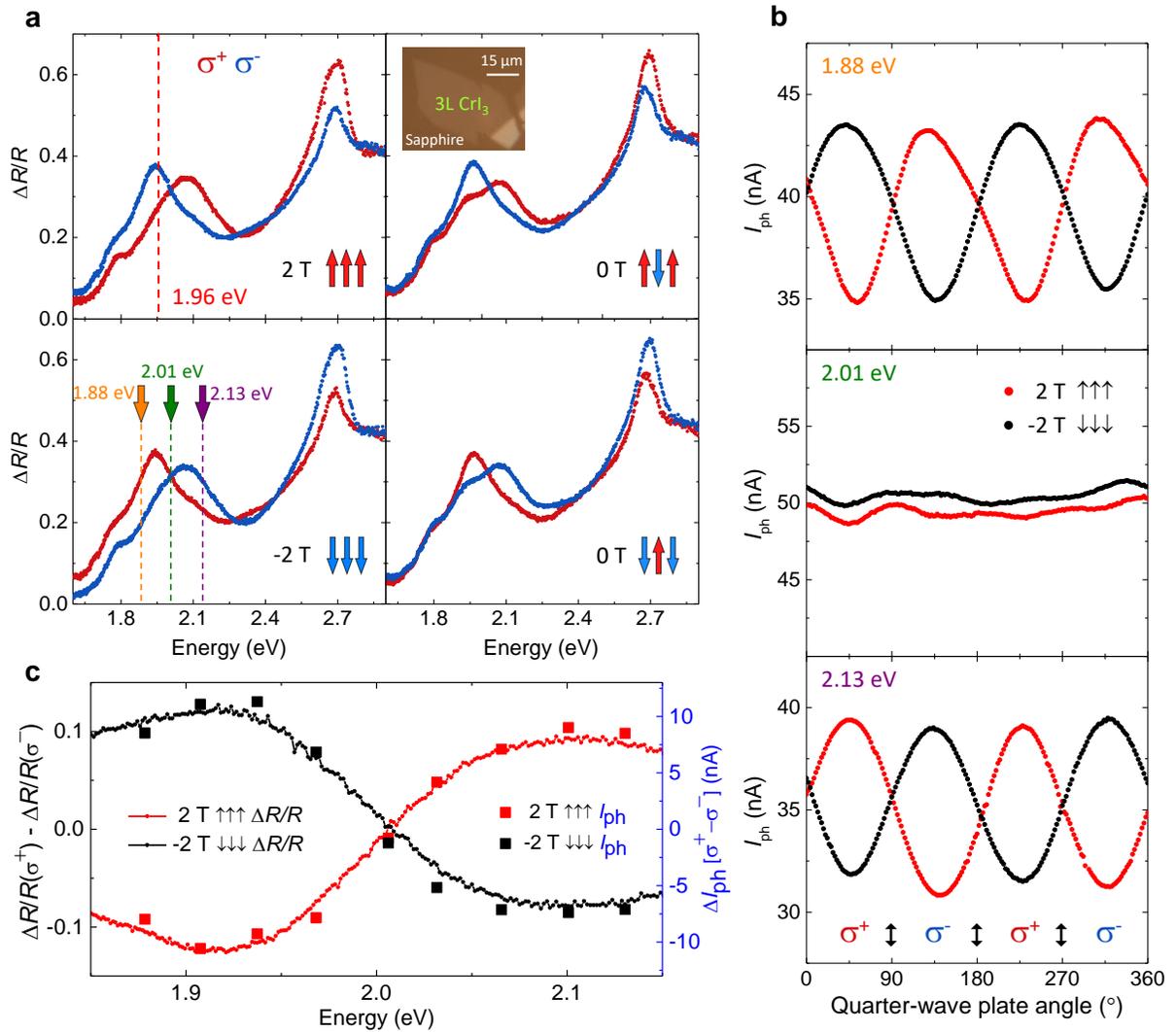

**Figure 4 | Interplay between magnetic order and photon helicity in absorption and photocurrent of 3L CrI$_3$. a**, Helicity-dependent $\Delta R/R$ spectra for all four magnetic states of 3L CrI$_3$ at selected magnetic fields. Red (blue) dots correspond to $\sigma^+$ ($\sigma^-$) photon helicity. Insets show the corresponding magnetic states and the optical microscopy image of a trilayer CrI$_3$ on sapphire. **b**, Photocurrent as a function of quarter-wave plate angle for ↑↑↑ state (2 T, red dots) and ↓↓↓ state (-2 T, black dots) measured with three selected photon energies indicated by the dashed lines in **a**. **c**, $\Delta R/R$ helicity difference (($\Delta R/R(\sigma^+)-\Delta R/R(\sigma^-)$, curve) and the overlaid change in photocurrent ($\Delta I_{ph}[\sigma^+-\sigma^-] = I_{ph}(\sigma^+)-I_{ph}(\sigma^-)$, squares) as a function of photon energy for ↑↑↑ state (2 T, red) and ↓↓↓ state (-2 T, black).

# Supplementary Information for

# Spin Photovoltaic Effect in Magnetic van der Waals Heterostructures

**Authors:** Tiancheng Song[1], Eric Anderson[1], Matisse Wei-Yuan Tu[2], Kyle Seyler[1], Takashi Taniguchi[3], Kenji Watanabe[4], Michael A. McGuire[5], Xiaosong Li[6], Ting Cao[6], Di Xiao[7], Wang Yao[8], Xiaodong Xu[1, 6]*

**Affiliations:**
[1]Department of Physics, University of Washington, Seattle, Washington 98195, USA.
[2]Department of Physics, National Cheng Kung University, Tainan 70101, Taiwan.
[3]International Center for Materials Nanoarchitectonics, National Institute for Materials Science, 1-1 Namiki, Tsukuba 305-0044, Japan.
[4]Research Center for Functional Materials, National Institute for Materials Science, 1-1 Namiki, Tsukuba 305-0044, Japan.
[5]Materials Science and Technology Division, Oak Ridge National Laboratory, Oak Ridge, Tennessee 37831, USA.
[6]Department of Materials Science and Engineering, University of Washington, Seattle, Washington 98195, USA.
[7]Department of Physics, Carnegie Mellon University, Pittsburgh, Pennsylvania 15213, USA.
[8]Department of Physics, University of Hong Kong, Hong Kong, China.

*Correspondence to: xuxd@uw.edu

**Contents:**

**Supplementary Text S1**
**Supplementary Table S1**
**Supplementary Figures S1 to S5**
**Fig. S1: Power dependence of CrI$_3$ photocurrent.**
**Fig. S2: Full Dependence of photocurrent on magnetic order in four-layer CrI$_3$.**
**Fig. S3: Photocurrent mapping in four-layer CrI$_3$.**
**Fig. S4: Helicity-dependent differential reflectance ($\Delta R/R$) for trilayer CrI$_3$ at 100 K.**
**Fig. S5: Helicity-dependent differential reflectance ($\Delta R/R$) for six-layer CrI$_3$.**

**Supplementary Text S1: Magnetic-state-dependent photocurrent model**

The photocurrent originates from the asymmetric potential between the top junction and bottom graphene-CrI$_3$ junctions. This asymmetry can arise from the different initial doping of graphene sheets and the built-in electric field induced by the device fabrication process. Moreover, the magnetic states of CrI$_3$ can also induce an intrinsic contribution to the contact junction asymmetry, which could potentially explain the magnetic-state-dependent photocurrent shown in Fig. 2a for a four-layer CrI$_3$. In the photocurrent device, the different magnetic states of CrI$_3$ result in different wavefunctions of the magnetic sub-bands, thus affecting the junction asymmetry. We illustrate this intrinsic mechanism with the following simplified model.

Due to the strong vertical confinement of each magnetic layer, the multilayer $CrI_3$ can be modelled as coupled magnetic quantum wells, the tunneling between which conserves in-plane momentum. At a given in-plane momentum, the Hamiltonian describing the coupling between the quantum well states if spin $\sigma$ simply reads $H_\sigma = \sum_{n=1}^{N_l} \varepsilon_{\sigma n} d_{\sigma n}^+ d_{\sigma n} + t(\sum_{n=1}^{N_l-1} d_{\sigma n}^+ d_{\sigma n+1} + h.c.)$, where $t$ is the interlayer hopping. Here we take the number of layers $N_l = 4$ to study the four-layer $CrI_3$ magnetic-state-dependent photocurrent. The magnetic states can be encoded in $\varepsilon_{\sigma n} = \frac{\sigma \Delta_0}{2} S_n - (n-1)\delta\varepsilon$, where $S_n = \pm 1$ is arranged to represent the magnetization of each layer, $\Delta_0$ is the ferromagnetic spin-split gap, and $\delta\varepsilon = eEd$ is the voltage drop caused by the asymmetric doping induced electric field $E$ over the inter-layer spacing $d$.

We then straightforwardly diagonalize this 4x4 Hamiltonian for $\sigma = \downarrow$ explicitly,

$$H_\downarrow = \begin{pmatrix} -\frac{\Delta_0}{2}S_1 & t & 0 & 0 \\ t & -\frac{\Delta_0}{2}S_2 - \delta\varepsilon & t & 0 \\ 0 & t & -\frac{\Delta_0}{2}S_3 - 2\delta\varepsilon & t \\ 0 & 0 & t & -\frac{\Delta_0}{2}S_4 - 3\delta\varepsilon \end{pmatrix}.$$

The eigenvalue equation is given by $H_\downarrow |u\rangle = E \sum_{n=1}^{N_l} a_n |n\rangle$. We then take the eigenvector corresponding to the lowest-energy eigenvalue, and obtain the components' amplitudes, $|a_n|^2$. The asymmetry can be simply characterized by the nonzero value of $|a_1|^2 - |a_4|^2$, which depends directly on the spin configuration $\{S_1, S_2, S_3, S_4\}$. We demonstrate this asymmetry with reasonable choices for the parameters, $\Delta_0 = 0.5$ eV, $t = 0.1$ eV, and $\delta\varepsilon = 0.025$ eV. The resulting wave amplitudes for $\sigma = \downarrow$ are shown in the Supplementary Table S1. In general, the weight of the wavefunction on the first layer (in contact with the top graphene electrode) is different from that on the fourth layer (in contact with the bottom graphene electrode), demonstrating the magnetic-state-dependent asymmetry. However, we note that the calculated wavefunction depends on the chosen parameters sensitively, and these cannot be uniquely extracted from experimental results. The precise determination of these parameters to quantify the asymmetry requires further experimental and theoretical studies.

| $\{S_1, S_2, S_3, S_4\}$ | $|a_1|^2$ | $|a_2|^2$ | $|a_3|^2$ | $|a_4|^2$ | $|a_1|^2 - |a_4|^2$ |
|---|---|---|---|---|---|
| $\{+,-,+,-\}$ | 0.0504 | 0.0419 | 0.8749 | 0.0328 | 0.0176 |
| $\{+,+,-,+\}$ | 0.3190 | 0.5262 | 0.0344 | 0.1204 | 0.1986 |
| $\{+,-,+,+\}$ | 0.0047 | 0.0136 | 0.4713 | 0.5104 | -0.5057 |
| $\{+,+,+,+\}$ | 0.0607 | 0.2531 | 0.4291 | 0.2571 | -0.1964 |

**Table S1 | Magnetic-state-dependent wavefunction of four-layer CrI$_3$.**

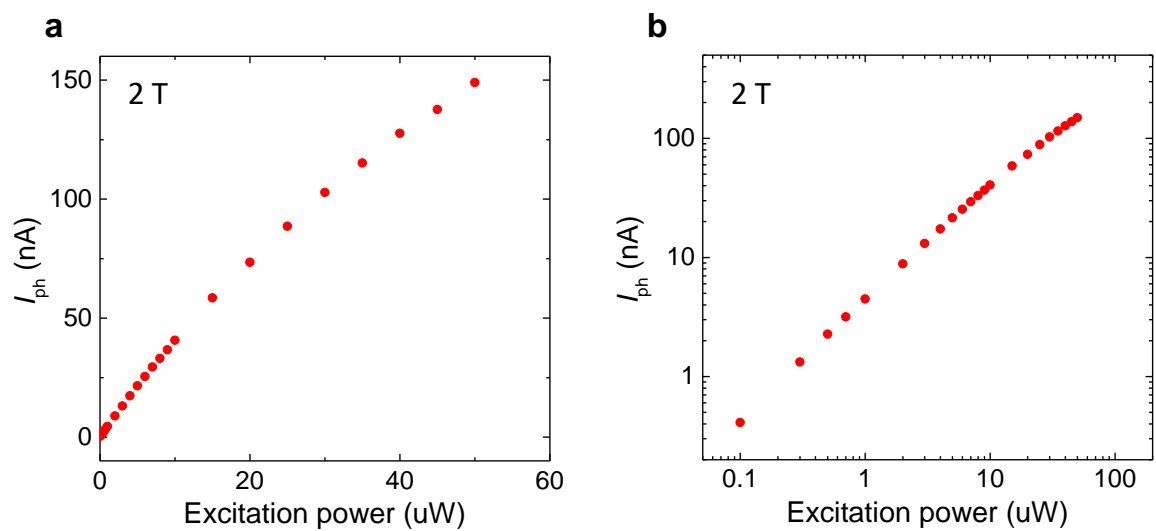

**Figure S1 | Power dependence of CrI$_3$ photocurrent. a**, **b**, Photocurrent as a function of the laser excitation power measured from the trilayer CrI$_3$ junction device (D1) plotted with linear and semi-log scales, respectively.

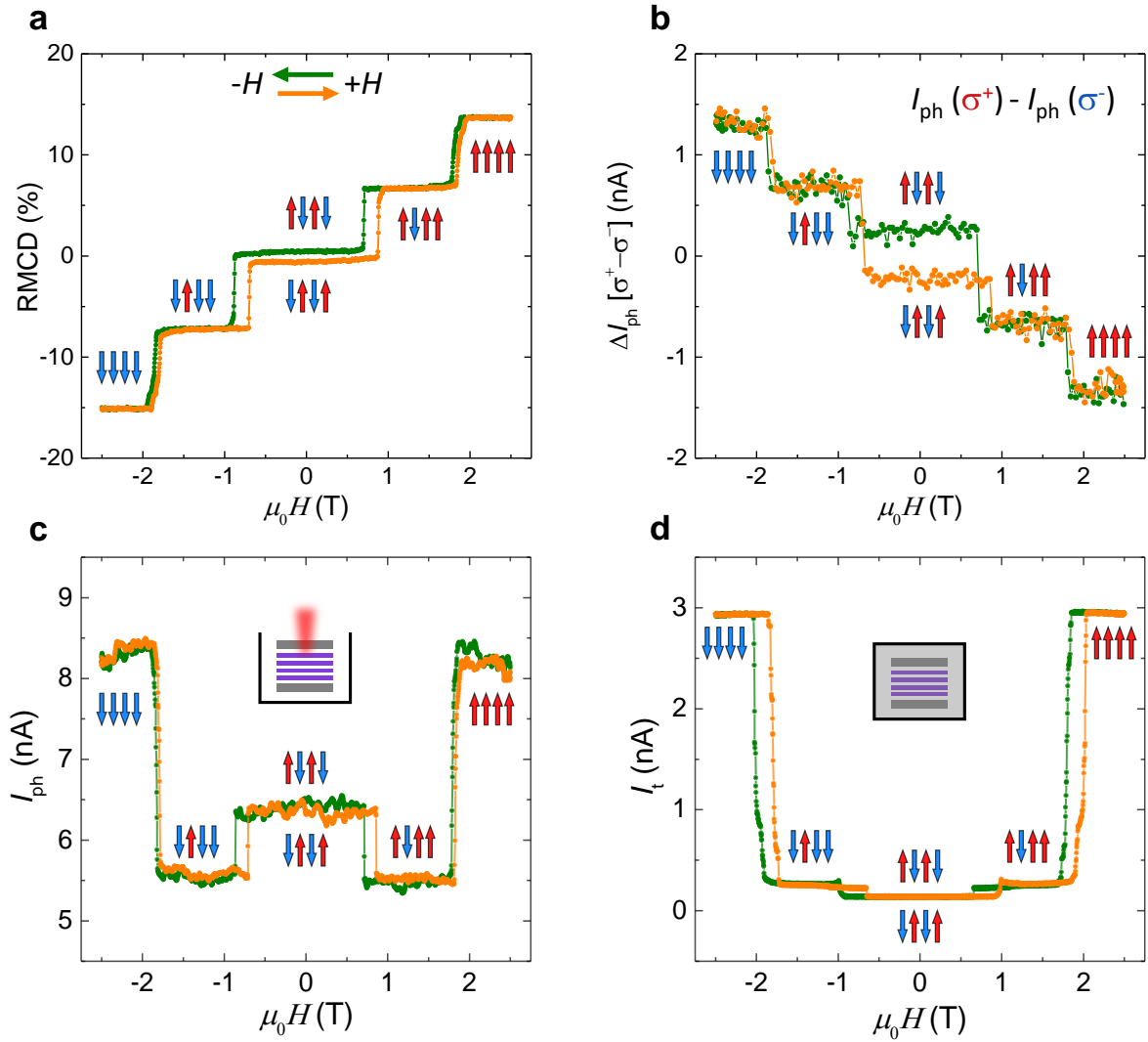

**Figure S2 | Full Dependence of photocurrent on magnetic order in four-layer CrI$_3$. a**, RMCD as a function of $\mu_0H$ for the four-layer CrI$_3$ junction device (D2). Insets show the corresponding magnetic states. **b**, The change in photocurrent ($\Delta I_{ph}$ [$\sigma^+$-$\sigma^-$] = $I_{ph}(\sigma^+)$-$I_{ph}(\sigma^-)$) as a function of $\mu_0H$ measured with an optical power of 1 µW. The net $\Delta I_{ph}$ for the AFM ground states at zero field can be attributed to the asymmetry in the four-layer CrI$_3$ flake, which also manifests as the finite nonzero RMCD signal. **c**, Photocurrent as a function of $\mu_0H$ measured with an optical power of 1 µW. **d**, Tunneling current ($I_t$) as a function of $\mu_0H$ measured from the same device at 80 mV bias in dark condition. Insets are the schematics of the device.

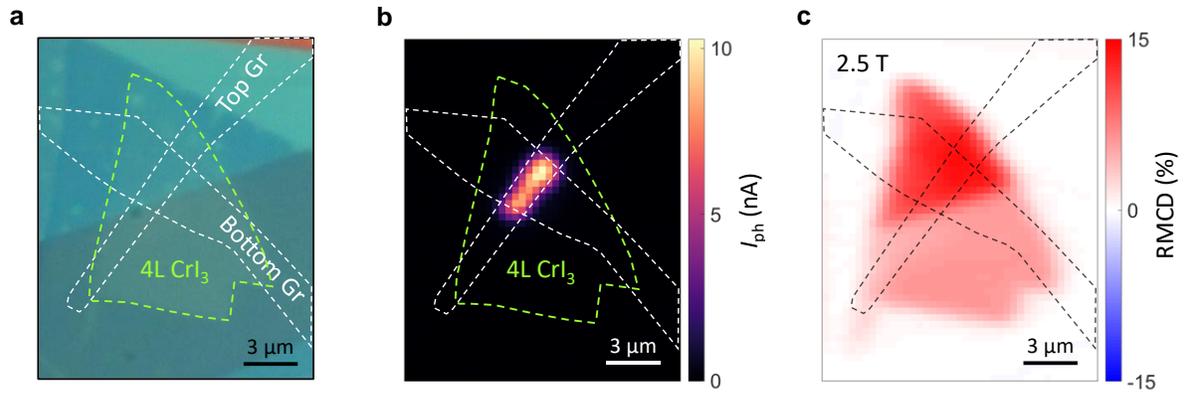

**Figure S3 | Photocurrent mapping in four-layer CrI$_3$. a**, Optical microscopy image of the four-layer CrI$_3$ junction device (D2) (scale bar, 3 µm). **b**, **c**, Spatial maps of photocurrent and RMCD signal measured from the same device at 2.5 T with an optical power of 1 µW (scale bar, 3 µm).

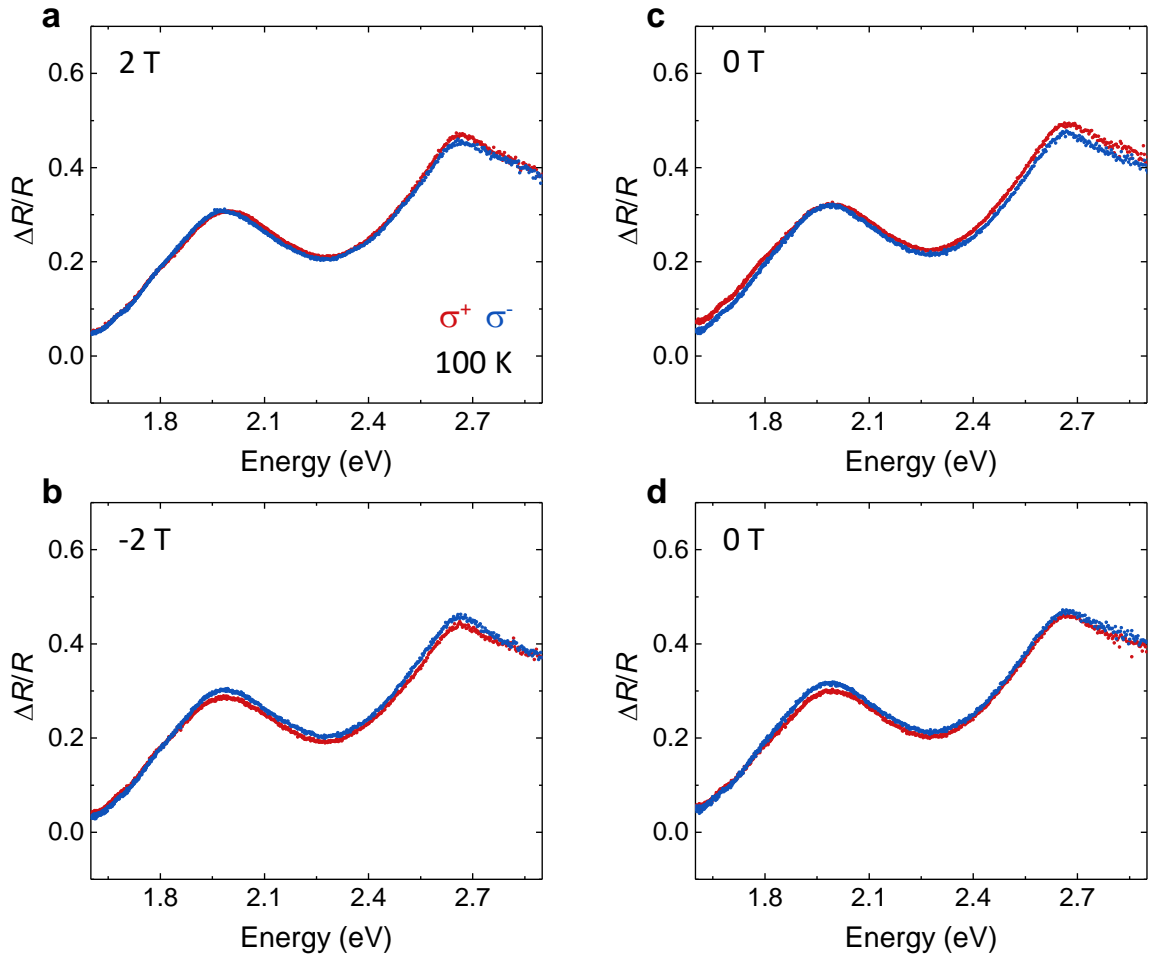

**Figure S4 | Helicity-dependent differential reflectance (ΔR/R) for trilayer CrI$_3$ at 100 K.**
**a-d**, Helicity-dependent ΔR/R spectra measured from the same trilayer CrI$_3$ on sapphire shown in Fig. 4a at the same selected magnetic fields. Red (blue) curve corresponds to σ$^+$ (σ$^-$) photon helicity.

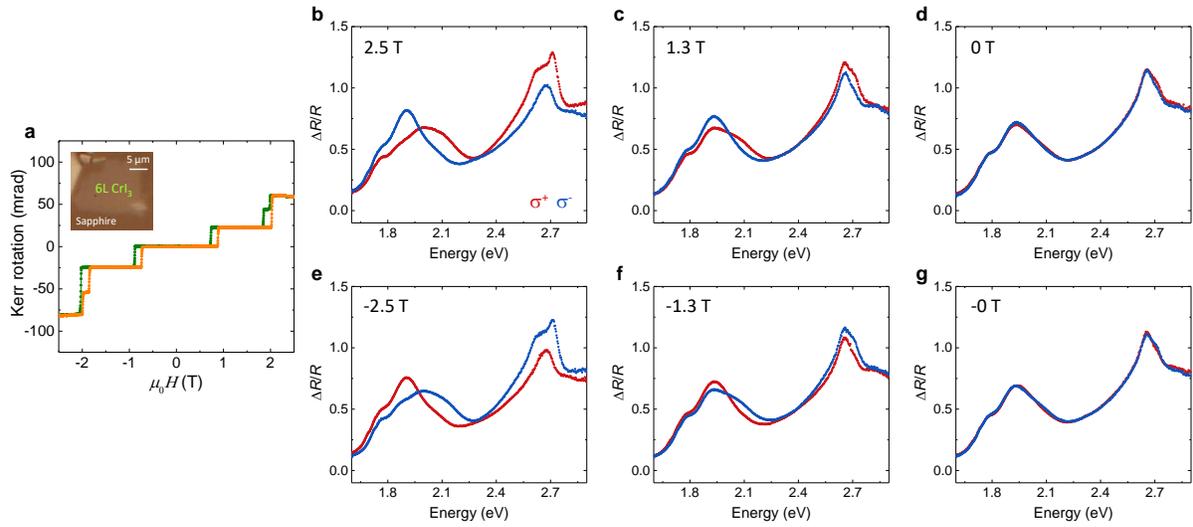

**Figure S5 | Helicity-dependent differential reflectance ($\Delta R/R$) for six-layer CrI$_3$. a**, Kerr rotation as a function of $\mu_0 H$ measured from a six-layer CrI$_3$ on sapphire at 2 K. Green (orange) curve corresponds to decreasing (increasing) magnetic field. **b-d**, Helicity-dependent $\Delta R/R$ spectra as magnetic field decreases from 2.5 T to 0 T. **e-g**, Helicity-dependent $\Delta R/R$ spectra as magnetic field increases from -2.5 T to 0 T. Red (blue) curve corresponds to $\sigma^+$ ($\sigma^-$) photon helicity.